\documentclass[journal=jctcce,manuscript=article,layout=traditional]{achemso}
\usepackage{helvet}
\usepackage{xcolor}
\usepackage{booktabs}
\usepackage{gensymb}
\usepackage{siunitx}

\usepackage[utf8]{inputenc}
\usepackage{url}

\title{Reverse translocation of a nascent polypeptide through the ribosomal exit tunnels}

\author{Felipe C. Nepomuceno}
\affiliation{Department of Physical Chemistry, University of Chemistry and Technology, Technická 5, Prague, Czech Republic}

\author{Petr Chalupský}
\affiliation{Department of Physical Chemistry, University of Chemistry and Technology, Technická 5, Prague, Czech Republic}

\author{Michal Kolář}
\affiliation{Department of Physical Chemistry, University of Chemistry and Technology, Technická 5, Prague, Czech Republic}
\email{kolarh@vscht.cz}

\begin{document}

\singlespacing

\maketitle

\begin{abstract}
Before they mature, all known proteins translocate through the ribosomal exit tunnel in a form of extended or partially folded nascent polypeptide. This translocation occurs in the natural direction from the peptidyl transferase centre buried deep in the large ribosomal subunit through a tunnel to the ribosomal surface. Some proteins, however, enter the ribosomal tunnel from outside and translocate in the reverse direction. In this work, we address a simple question: to what extent is the ribosomal tunnel intrinsically directional. To compare the forward and reverse peptide translocations, we performed biased molecular dynamics simulations and assessed the directionality of tunnels from four different organisms using a model poly-alanine decapeptide. Our simulations reveal that the tunnel architecture intrinsically favours the natural direction translocation from peptidyl transferase centre to the ribosomal surface. Consequently, proteins that occupy the tunnel in the reverse direction likely require dedicated structural machinery to overcome this intrinsic bias.
\end{abstract}

\section{Introduction}

Ribosomes are the sole biomolecular nanomachines responsible for protein synthesis. During elongation, nascent polypeptide (NP) exits the peptidyl transferase centre (PTC) through the ribosomal tunnel, a void approximately \SI{10}{\nano\meter} long and 0.8--\SI{1.5}{\nano\meter} in diameter that traverses the large ribosomal subunit \cite{Voss2006a, DaoDuc2019}. The tunnel walls are primarily composed of ribosomal RNA (rRNA), with contributions from several ribosomal proteins, most notably uL4 and uL22 at the constriction site and uL23 and uL24 in the vestibule \cite{Brodersen2005, Wilson2005}. The natural direction of NP translocation through the exit tunnel is from the PTC toward the ribosome surface. As translation progresses, the N-terminus of the NP moves through the tunnel first. The C-terminus remains covalently attached to the P-site tRNA and translocates through the tunnel only after the translation termination by a release factor \cite{Samatova2024,Kolar2024}.

Exceptions to this unidirectional translocation have been reported, however. In mammalian mitochondrial ribosomes, for example, the mitochondria-specific protein mL45 positions its N-terminal extension in the exit tunnel from the vestibule side, reducing the space available to the NP that limits helix formation \cite{Itoh2021}. The mL45 N-terminus is displaced as the nascent polypeptide grows, implying a transient reverse occupancy of the tunnel lumen \cite{Kummer2018}. In eukaryotic cytosolic ribosomes, a conceptually similar reverse insertion is carried out by the nascent polypeptide-associated complex (NAC), an abundant heterodimeric chaperone composed of the NAC$\alpha$ and NAC$\beta$ subunits \cite{Lee2026}. NAC binds to the ribosome surface near the tunnel exit with nanomolar affinity \cite{Jomaa2022} and plays a central role in coordinating the access of protein biogenesis factors to emerging NPs \cite{Wiedmann1994,Gamerdinger2019}. Remarkably, NAC inserts the N-terminal tail of its $\beta$-subunit (N-NAC$\beta$) into the ribosome exit tunnel in the direction opposite to nascent chain translocation \cite{Gamerdinger2019}. In this conformation, NAC can sense the growing nascent chain close to the PTC and block premature access of other factors such as those targetting the NP to the Sec61 translocon. As elongation proceeds, the N-NAC$\beta$ segment is displaced from the tunnel interior by the growing polypeptide and relocates to a canonical binding site on the ribosome surface \cite{Gamerdinger2019}.

These examples show that the exit tunnel is not a purely one-way conduit dedicated to outward NP movement, but rather a channel capable of accommodating occupancy and transit in either direction, at least for accessory elements such as the mL45 extension and N-NAC$\beta$. This raises the question of whether the tunnel's bidirectional character extends to translocation of the NP itself, and whether the geometry and chemical composition of the tunnel walls impose an intrinsic directional preference on such movement. To date, this question of tunnel directionality has not been directly addressed even though intuition might suggest the answer is obvious.

Structural biology techniques, while indispensable for resolving the architecture of the tunnel and the binding poses of factors such as mL45 and NAC, are not well suited to answering this question on their own. They excel at describing the end states of a peptide translocation event but structural-biology methods resolving the continuous trajectories or directional dynamics that connect them are only emerging \cite{Frank2017, Klebl2023}. The problem is, however, ideal for atomistic molecular dynamics simulations in the non-equilibrium regime.

Molecular dynamics (MD) simulations have proven invaluable for studying NP dynamics within the ribosomal exit tunnel \cite{Bock2023}. Coarse-grained and all-atom simulations have been used to characterize NP escape after translation termination \cite{Bui2016, Bui2018, Bui2021, Bui2023, Nissley2020}, the folding of secondary structure elements within the tunnel \cite{OBrien2008, Bano-Polo2018, Kolar2022}, and the interactions of arresting peptides with the tunnel walls \cite{Gumbart2012, Arenz2016b}. Force-probe MD (fpMD) simulations, in particular \cite{Grubmuller1996, Rief2002}, have enabled the study of NP translocation under controlled non-equilibrium conditions. In a recent study, we used fpMD to investigate the effect of pulling versus pushing on peptide conformational dynamics in carbon nanotubes and found that the amino acid sequence plays a more important role than the directionality of force application \cite{Nepomuceno2025}. Under simulation conditions, only minor differences were observed between pulling the peptide and pushing it in the same direction. Extending such simulations to realistic ribosome exit tunnels would provide direct insight into whether the tunnel itself imposes directional preferences on translocation.

Here, we use fpMD simulations to investigate the directionality of polypeptide translocation through ribosome exit tunnels from four organisms: \textit{Escherichia coli}, \textit{Saccharomyces cerevisiae}, \textit{Homo sapiens} (cytosolic), and \textit{Homo sapiens} (mitochondrial). For each tunnel, we performed pulling simulations of a model alanine decapeptide in both the natural direction (from the PTC toward the tunnel exit, Fig.\,\ref{fig:overview}A) and the reverse direction (from the tunnel exit toward the PTC, Fig.\,\ref{fig:overview}B). Our results reveal how the geometry and chemical heterogeneity of exit tunnels from the three domains of life and from mitochondria affect the directionality of molecular transport through these confined biological channels.

\section{Methods}

\subsection{Construction of ribosomal exit tunnel models}

Four ribosomes were studied that are summarised in Tab.\,\ref{tab:systems}: \textit{E.\,coli} (EC), \textit{S.\,cerevisiae} (SC), \textit{H.\,sapiens} cytosolic (HS), and \textit{H.\,sapiens} mitochondrial (MT). The exit tunnels of these ribosomes were extracted using the following procedure.

\begin{table}[tb]
    \centering
    \begin{tabular}{lllllll}
    \toprule
    abbr. & organism & PDB & resolution [\AA] & trajs & ns/traj \\
    \midrule
    EC &  \emph{E. coli} & 7K00\cite{Watson2020} & 1.98 & 2x16 & 200 \\
    SC &  \emph{S. cerevisiae} & 7TOO\cite{Loveland2022} & 2.70 & 2x16 & 200\\
    HS &  \emph{H. sapiens} (cytosolic) & 8A3D\cite{Faille2023} & 1.67 & 2x16 & 200 \\
    MT &  \emph{H. sapiens} (mitochondrial) & 7QI4\cite{Singh2024} & 2.21 & 2x16 & 200 \\
    \bottomrule
    \end{tabular}
    \caption{Overview of the simulated systems.}
    \label{tab:systems}
\end{table}

The tunnel axis was defined as a vector connecting the centres of mass of the PTC residues and the exit port residues. To capture the curvature of the tunnel, which is poorly described by the straight axis, a series of local centres of mass was computed: for each \SI{0.1}{\nano\meter}-thick layer of residues lying within a \SI{1.8}{\nano\meter} radius around the axis, the centre of mass was calculated. The resulting set of centres of mass defined the curved tunnel path.

All residues within \SI{1.8}{\nano\meter} of any of these centres of mass were included in the tunnel selection. Where visual inspection revealed a hole in the tunnel wall, this radius was increased up to \SI{2.2}{\nano\meter} until the hole was plugged. Non-canonical terminal residues, such as pseudouridine, were removed, as were isolated single-residue fragments. No tRNAs were retained at the PTC.

The resulting tunnel models contained between 10{,}000 and 20{,}000 non-hydrogen atoms and resembled cylinders with a base diameter of approximately \SI{7}{\nano\meter} and a height of approximately \SI{10}{\nano\meter}.

\begin{figure}[tb]
    \centering
    \includegraphics{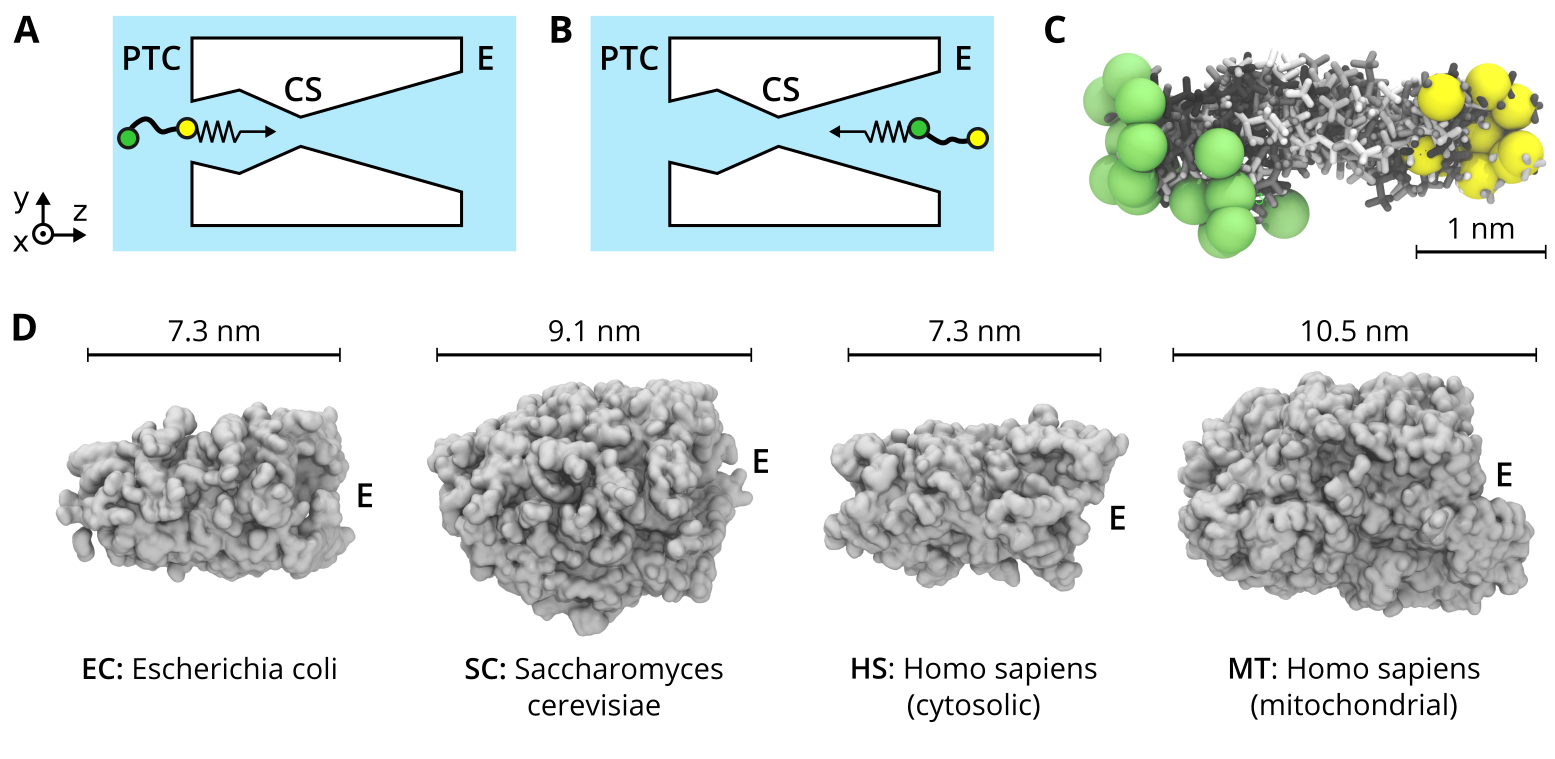}
    \caption{Overview of the simulated systems. A) Scheme of natural-direction translocation from the peptidyl transferase centre (PTC) through the constriction site (CS) to the tunnel exit (E). B) Scheme of opposite-direction translocation. C) Diversity of initial peptide conformations, with the N-terminal acetyl group shown in yellow and the C-terminal N-methylamide group shown in green. D) Surface representations of the four tunnels investigated, with the exit marked E; the tunnel lengths used for normalisation are indicated.}
    \label{fig:overview}
\end{figure}

\subsection{System setup}

Each tunnel was placed in a tight rectangular periodic box, ensuring a minimum distance of \SI{0.3}{\nano\meter} between the solute and the periodic boundary. All non-hydrogen atoms were subjected to harmonic position restraints with a force constant of \SI{300}{\kilo\joule\per\mole\per\nano\meter\squared} to prevent large conformational changes while permitting local fluctuations. The box was filled with SPC/E water \cite{Berendsen1987} and neutralized with K\textsuperscript{+} ions using the Joung and Cheatham parametrization \cite{Joung2008}.

Interactions were described using the Amber family of force fields: ff10 for canonical RNA \cite{Cornell1995,Perez2007,Zgarbova2011}, ff12SB for protein~\cite{Maier2015}, and the parameters of Aduri \textit{et\,al.} for non-canonical RNA \cite{Aduri2007}. This combination of force fields has been proved reliable for our previous simulations of the ribosome \cite{McGrath2022,Kolar2022} as well as in studies of others \cite{Huter2017,Beckert2021}.

All simulations were done in GROMACS 2019 \cite{Abraham2015}. The electrostatics was treated using particle mesh Ewald methods with a direct space cut-off of \SI{1.2}{\nano\meter} and Fourier spacing of \SI{0.12}{\nano\meter}. Van der Waals interactions were modelled by a switched Lennard-Jones potential with the cut-off of \SI{1.2}{\nano\meter} and the switch applied beyond \SI{1.0}{\nano\meter}. Bond lengths of bonds involving hydrogens were constrained using P-LINCS \cite{Hess2008a}. The production simulations were done at NVT ensemble at 310 K using v-rescale thermostat \cite{Bussi2007}.

\subsection{Force-probe simulations of deca-alanine}

Deca-alanine was prepared in the fully extended conformation and capped with an acetyl group (ACE) at the N-terminus and an N-methylamide group (NME) at the C-terminus. Its conformations were sampled in a \SI{1}{\micro\second} NPT simulation at \SI{310}{\kelvin} and \SI{1}{\bar}. From the resulting ensemble, we selected 16 conformations to be placed near the PTC or the exit port of the tunnel (Fig.\,\ref{fig:overview}C). Only sufficiently extended conformations were considered, using the same criterion as in our previous study of deca-peptides in CNTs \cite{Nepomuceno2025}. Namely, a peptide was deemed extended only if neither terminus was bent such that a terminal residue extended beyond the third residue.

The placement of each peptide was performed in an automated manner. A reference fully extended peptide was placed manually so as to approximately align with the tunnel axis, with the N-terminus near the tunnel end, and each of the 16 conformations was then superimposed onto this reference using the C$_\alpha$ atoms.

In this way, two sets of fpMD simulations were generated for each tunnel, pulling the peptide either in the natural direction, from the PTC towards the tunnel exit, or in the opposite direction, from the exit towards the PTC, as depicted in Fig.\,\ref{fig:overview}AB. The orientation of the peptide reflected the natural orientation of the nascent peptide: the N-terminus of deca-alanine was pulled towards the tunnel exit, and the C-terminus towards the PTC.

Prior to the pulling simulations, each system was equilibrated in a series of short simulations. The peptide positions were restrained by a harmonic potential with a force constant of \SI{1000}{\kilo\joule\per\mole\per\nano\meter\squared}. First, the system's potential energy was minimized using approximately 10{,}000 steepest-descent steps, or until the maximum force fell below \SI{100}{\kilo\joule\per\mole\per\nano\meter}. Next, a \SI{1}{\nano\second} NVT simulation at \SI{300}{\kelvin} equilibrated the temperature of the solvent (water and ions) using the v-rescale thermostat, while the solute (tunnel and peptide) was kept at \SI{10}{\kelvin}. Velocities were drawn randomly from the Maxwell--Boltzmann distribution at \SI{5}{\kelvin}. A subsequent \SI{1}{\nano\second} NPT simulation at \SI{300}{\kelvin} and \SI{1}{\bar} equilibrated the box dimensions using Berendsen barostat. Finally, a \SI{2}{\nano\second} NVT simulation at \SI{310}{\kelvin} equilibrated the temperature of the entire system.

The fpMD simulations were performed in the constant-velocity regime: the force probe was attached to the centres of mass of the terminal ACE or NME groups and moved along the tunnel at a speed of \SI{0.05}{\nano\meter\per\nano\second} with a force constant of \SI{602.7}{\kilo\joule\per\mole\per\nano\meter\squared}, using the centre of mass of the tunnel as the reference group. The resulting loading rate, defined as the product of the force-probe speed and the force constant, was \SI{0.1}{\pico\newton\per\pico\second}. This loading rate was found to be optimal in our previous simulations of deca-peptides in CNTs~\cite{Nepomuceno2025}, reflecting a trade-off between the speed of peptide translocation and the timescales achievable with finite computational resources. Only the Z Cartesian coordinate of the probed residue was biased, so the deca-alanine was free to move in the XY plane. Each fpMD simulation lasted \SI{200}{\nano\second}, over which the force probe moved by \SI{10}{\nano\meter}.

\subsection{Analyses}

The progress of the peptide through the tunnel over the course of the trajectories was monitored using the Z coordinate of the probed residue. Because the tunnels were aligned with the Z-axis, this provided information about the relative position of the peptide with respect to the tunnel.

The total displacement of the peptide was defined as the difference between the final and initial positions of the probed residue. Each tunnel had a different length, measured as the difference in the Z-component between the residues at the PTC and at the tunnel exit. We therefore introduced the relative displacement in \%, defined as the total displacement normalized by the tunnel length. Unless stated otherwise, replicas in which the relative displacement did not reach 80\% during the simulation time were excluded from the analysis.

The force analysis was based on the harmonic forces output by GROMACS as a function of simulation time, sampled every \SI{0.1}{\pico\second}.

The solvation status of the peptide was assessed from the number of water molecules within \SI{0.35}{\nano\meter} of any peptide atom.

\section{Results and discussion}

\subsection{Ribosomal tunnel is intrinsically directional}

To test whether the ribosomal exit tunnel imposes a directional bias on nascent polypeptide translocation, we performed fpMD simulations in which a probed residue was pulled through the tunnel in both the natural (forward) direction and the reverse direction. For each of the four tunnels examined, we generated 16 independent replicas per direction. The replicas differed in the conformation of the peptide and the initial velocities.

In each simulation, the force probe was displaced by \SI{10}{\nano\meter} over the course of a \SI{200}{\nano\second} simulation. However, because the probed residue is mechanically coupled to the probe through a compliant spring, its actual displacement can lag substantially behind that of the probe whenever it encounters steric or interaction resistance from the tunnel walls. The total displacement of the probed residue therefore serves as a direct readout of the resistance it experiences along its path, and, by extension, as a coarse-grained descriptor of the internal geometry and directionality of each tunnel.

\begin{figure}[tb]
    \centering
    \includegraphics{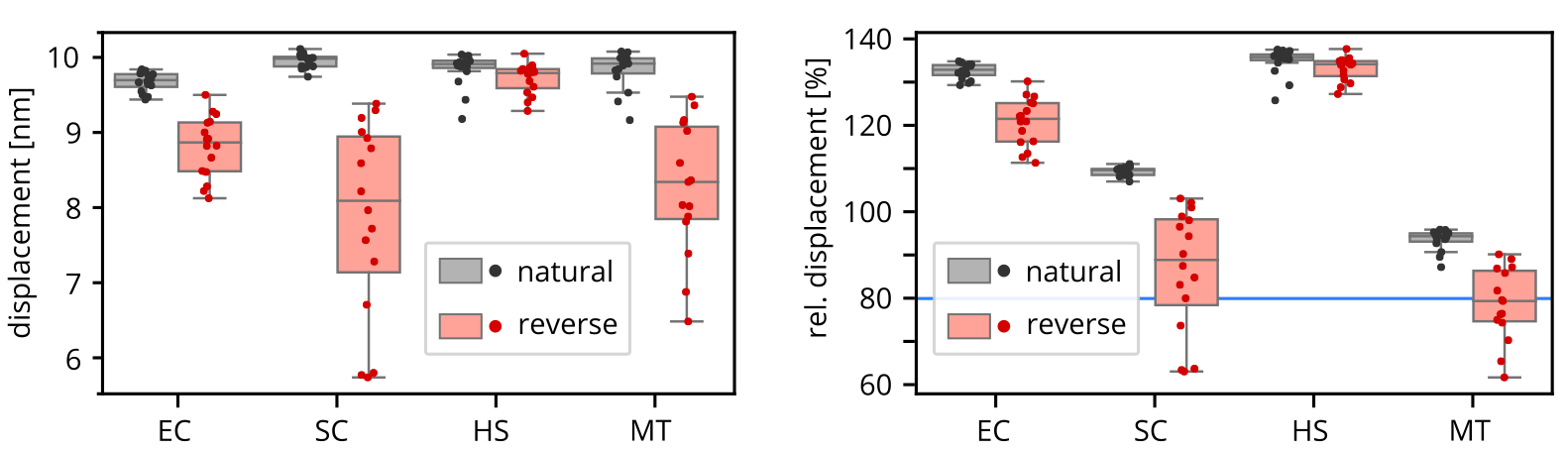}
    \caption{The absolute (left) and relative (right) displacement of the probed residue  in the natural (from PTC) and reverse (from tunnel exit) directions after the \SI{200}{\nano\second} fpMD simulations. The boxes show the range between first and third quartiles, the whiskers extend to the point that is 1.5 times interquartile range from the box. Medians are shown within the boxes. Individual points represent 16 independent fpMD trajectories. The blue line marks the displacement threshold used for filtering out the unsuccessful translocations.}
    \label{fig:displacement}
\end{figure}

Fig.\,\ref{fig:displacement} summarizes the total displacement of the probed residue after \SI{200}{\nano\second} of simulation, pooled across all replicas, tunnels, and both directions of translocation. Across all four tunnels, the total displacement ranged from approximately 6 to \SI{10}{\nano\meter}, depending on the tunnel identity and the direction in which the residue was pulled. 

In every tunnel, translocation in the natural direction resulted in greater net displacement of the probed residue than translocation in the reverse direction, indicating that the forward path is systematically less obstructed. The magnitude of this asymmetry, however, was not uniform across tunnels: reverse translocation was markedly less successful in the SC and MT tunnels than in the EC and HS tunnels, where the natural and reverse trajectories were more comparable in their extent of translocation. The reverse translocations in SC and MT tunnels were associated with substantially reduced displacement than in EC and HS.

Because the four tunnels differ in their absolute lengths, a direct comparison of raw displacement values can be confounded by tunnel geometry alone. To account for this, we normalized the total displacement of each trajectory by the corresponding tunnel length (Fig.\,\ref{fig:overview}) and expressed the result as a relative (percentage) displacement. 

A trajectory was classified as successful if the probed residue travelled more than 80\% of the tunnel length; trajectories that did not meet this criterion were considered unsuccessful and were excluded from further analysis. Using this normalized measure, we found that in all four tunnels the probed residue was able to traverse the entire tunnel in the natural forward direction, with relative displacements reaching or exceeding 100\%. This normalized analysis reinforces the conclusion drawn from the raw displacement data: reverse translocation was consistently more obstructed than forward translocation across all tunnels, and this obstruction was most pronounced in the SC and MT tunnels, where relative displacement in the reverse direction fell well short of the 80\% success threshold in a fraction of replicas.

\begin{figure}[tb]
    \centering
    \includegraphics{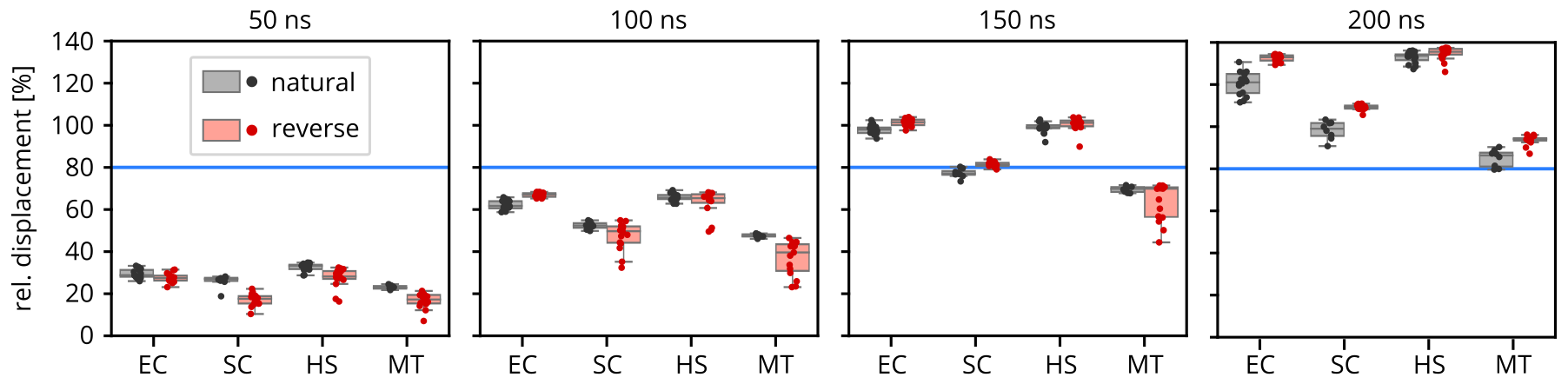}
    \caption{Relative displacement of the probed residue at four time points during the simulation. Boxes span the first to third quartiles, with the median indicated by the horizontal line inside each box; whiskers extend to the most extreme point within 1.5 times the interquartile range from the box. Individual points represent independent fpMD trajectories that resulted in successful translocation. The blue line marks the relative displacement threshold used to filter out unsuccessful translocations.}
    \label{fig:time}
\end{figure}

To gain further insight into the dynamics of translocation, we examined the relative displacement of the probed residue as a function of simulation time, considering only those trajectories that resulted in successful translocation as defined above. Fig.\,\ref{fig:time} shows the distribution of relative displacement across replicas at four representative time points during the simulation. 

A striking difference emerges between the two directions of translocation in terms of the spread of individual trajectories. In the natural forward direction, the probed residue advances at a consistent rate across nearly all replicas, resulting in a narrow distribution of relative displacement at each time point. In contrast, reverse translocations display considerable variability among replicas, with trajectories already beginning to diverge from one another within the first \SI{50}{\nano\second} of simulation. This broader spread indicates that, unlike the forward direction, the reverse pathway does not present a single dominant route through the tunnel, but rather a range of kinetically distinct outcomes, consistent with the presence of variable, direction-dependent barriers along the tunnel.

Taken together, these results demonstrate that the ribosomal exit tunnel possesses an intrinsic geometric and/or physicochemical bias that favours translocation in the natural direction, where the N-terminus heads toward the tunnel exit. This directional preference is observed across evolutionarily divergent ribosomes, from bacteria to eukaryotic cytosolic and mitochondrial systems, suggesting that it may be a conserved architectural feature of the tunnel rather than a species-specific adaptation.

\subsection{Force profiles}

\begin{figure}[tb]
    \centering
    \includegraphics[width=\textwidth]{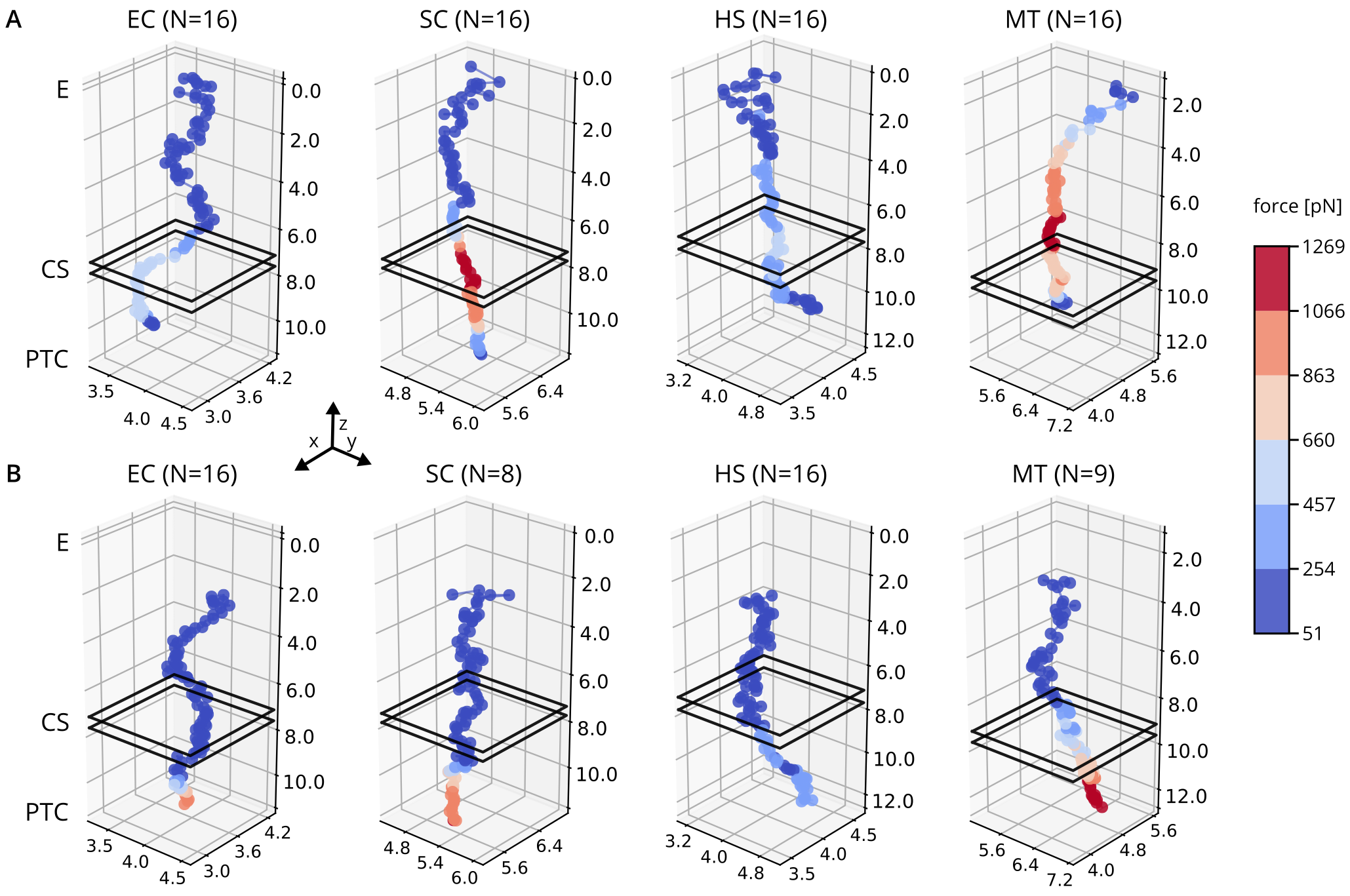}
    \caption{Trajectories of the probed residues along the tunnel, oriented with the peptidyl transferase centre (PTC) at the bottom and the tunnel exit (E) at the top. The constriction site (CS) region is shown as black squares. The colour scale goes from the low-force regime (blue) to the high-force regime (red). Trajectories represent averages over successful replicas, with the number of replicas $N$, indicated. Cartesian dimensions are in nanometres. A) Forward translocation from the PTC to the exit. B) Reverse translocation from the exit to the PTC.}
    \label{fig:forces}
\end{figure}

To characterise the energetics of translocation, we measured the force exerted by the force probe on the probed residue -- the N-terminal acetyl or the C-terminal N-methylamide capping group, depending on the direction of pulling. Fig.\,\ref{fig:forces} shows the resulting trajectories of the probed residues, averaged over the successful replicas and coloured according to the magnitude of the force itself, averaged over the same set of replicas; the number of replicas contributing to each average, $N$, is indicated in the figure. This representation allows the translocation pathway and its associated energetics to be visualised simultaneously, so that low-force and high-force regions can be identified for both directions of travel.

A first observation is that, in every tunnel and in both directions, the path traced by the probed residue is not straight. This lack of tunnel linearity is well appreciated in the existing literature \cite{Voss2006a,Kolar2024}. Furthermore it has repeatedly shown that the nascent chain may not progress along the central axis of the tunnel but instead weaves between the rRNA and protein walls that line it \cite{Ishida2008, Samatova2024}. Within our data, the nascent polypeptide appears to enjoy markedly greater lateral freedom while traversing the EC tunnel, in both the forward and the reverse direction, than while traversing any of the other tunnels considered here.

The CS stands out consistently as a region with elevated forces in all four tunnels. This is consistent with the view that the CS, formed by the extended loops of uL4 and uL22, imposes a substantial entropic penalty on the passing nascent chain by restricting the range of conformations it can adopt \cite{Petrone2008,McGrath2026}. The two directions of translocation, however, encounter this barrier at different stages of the process. During forward translocation the CS is reached early in the simulation, so that the associated high-force regime dominates the initial part of the trajectory. Reverse translocation follows the opposite pattern: it begins in a comparatively low-force regime and only slows down once the peptide approaches the CS, so that the dominant resistance is encountered towards the end of the simulation rather than at the outset.

A related and somewhat subtle point concerns the apparent timing of the high-force regime relative to the CS itself. Because the plotted trajectories track the position of a single terminal residue rather than the whole nascent chain, the high-force regime appears to lag behind the moment at which this residue has already cleared the constriction. In practice, the bulk of the nascent peptide remains lodged within the CS for some time after its leading residue has passed through, so the force experienced by the probe continues to reflect the resistance of the chain body still negotiating the constriction.

Comparing the four tunnels, translocation through the EC and HS tunnels is associated with systematically lower forces than translocation through the SC and MT tunnels, in both the forward and the reverse direction. This suggests that the energetic cost of translocation, and by extension the ease with which a nascent chain can be threaded through the tunnel, varies appreciably between the tunnel types studied.

The absolute magnitude of these forces should, however, be interpreted with caution. Our fpMD simulations span only \SI{200}{\nano\second}, whereas the residence time of a nascent-chain segment within a given region of the tunnel during physiological translation occurs on a timescale of tens to hundreds of milliseconds per elongation cycle. To induce translocation within a computationally tractable window, the probed residue must therefore be driven many orders of magnitude faster than it would move \emph{in vivo}, and the force required to do so is correspondingly inflated. This discrepancy is a well-documented feature of force-probe MD in general: pulling at rates far above experimental loading rates reliably yields forces well in excess of those measured at near-equilibrium rates \cite{Isralewitz2001,Rico2019}. 

We therefore expect the qualitative picture emerging from our simulations to hold at biological timescales, while the absolute forces reported here are almost certainly overestimates. This applies to both the directional asymmetry between forward and reverse translocation and the relative ranking of the four tunnels. Experimental measurements of the forces acting on nascent chains within the ribosome exit tunnel, obtained using arrest-peptide-based force sensors and optical-tweezer assays, place the physiologically relevant range at roughly 1--20\,pN \cite{Goldman2015,Nilsson2015a}, which is one to two orders of magnitude below the forces typically generated in nanosecond-timescale pulling simulations. We would accordingly expect the true forces experienced during ribosomal translocation to lie within, or close to, this range.

It should also be stressed that the reverse-translocation trajectories shown here represent only those replicas in which translocation was ultimately successful; replicas in which the peptide failed to reach the PTC are, by construction, excluded from the averages and therefore do not contribute to the depicted force profile.

Finally, the shape of the force profiles closely reflects the underlying geometry of the tunnel, as summarised in Fig.\,\ref{fig:overview}. The region between the PTC and the CS is considerably more confined than the region beyond the CS, where the tunnel gradually widens and its effective diameter increases towards the exit \cite{Voss2006a, Samatova2024, Yu2023a}. The correspondence between the geometric profile of the tunnel and the force experienced by the translocating peptide reinforces the interpretation that the high-force regions identified here arise primarily from steric and entropic restriction. Specific chemical interactions are expected minor in case of the chemically simple deca-alanine.

\subsection{Peptide solvation}

The ribosomal exit tunnel is filled with solvent molecules that surround the NP throughout translocation. Previous MD simulations showed that this confined solvent behaves very differently from bulk water \cite{Lucent2010b}. Rather than following the classical two-state picture of either a continuous dielectric fluid or discrete, tightly bound molecules, water inside the tunnel was found to be slow-diffusing and semi-structured, sampling a continuum of states between these two extremes. Such non-bulk-like behaviour has direct consequences for co-translational processes, including sequence-dependent ribosomal stalling, nascent-chain compaction, and folding within the tunnel \cite{Lucent2010b, Petrone2008}.

\begin{figure}[tb]
    \centering
    \includegraphics[width=\textwidth]{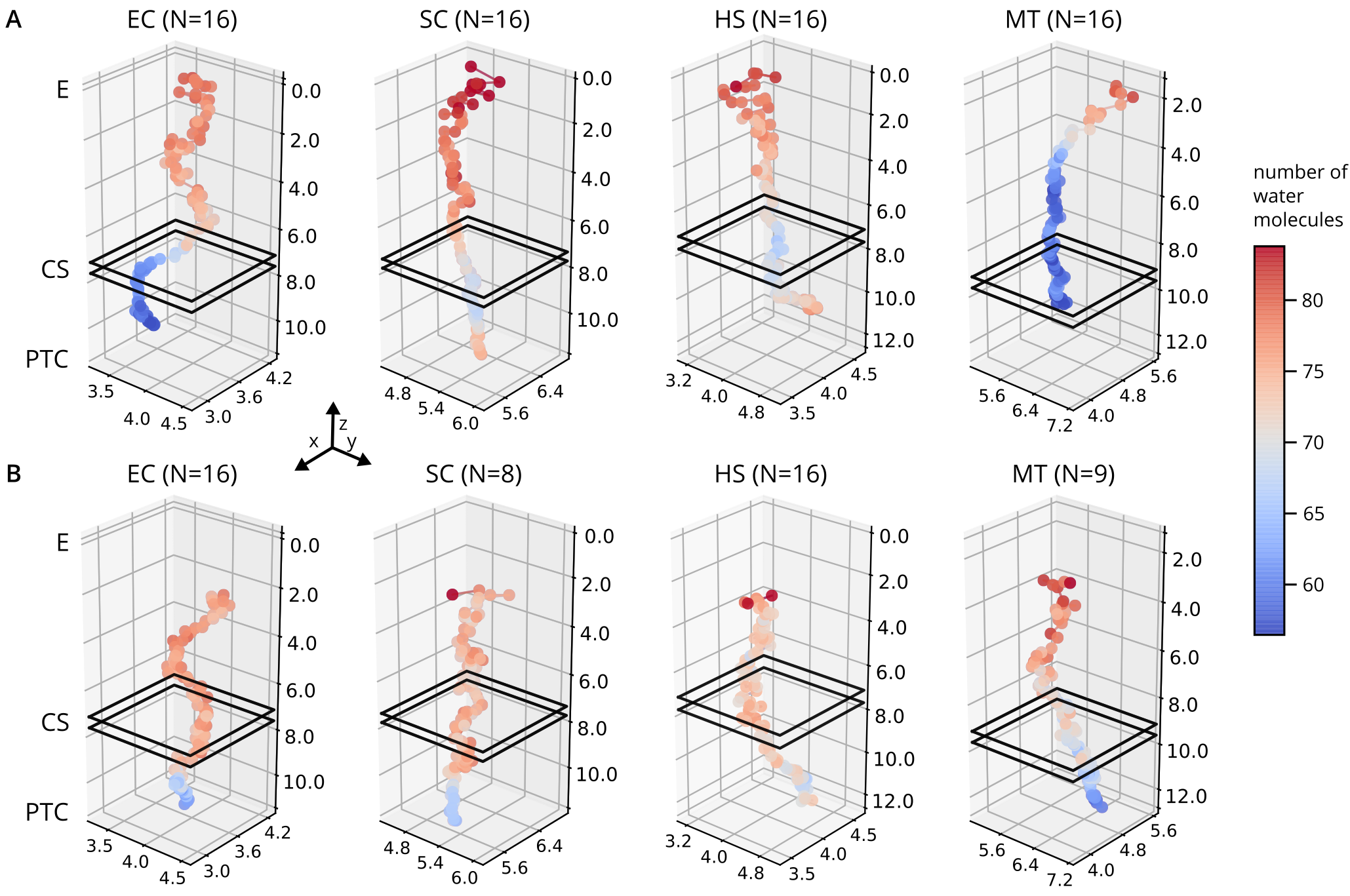}
    \caption{Trajectories of the probed residues along the tunnel, oriented with the peptidyl transferase centre (PTC) at the bottom and the tunnel exit (E) at the top. The constriction site (CS) region is shown as black squares. The colour scale represents the number of water molecules within \SI{0.35}{\nano\meter} from the peptide and it goes from the few waters (blue) to the many waters (red). Trajectories represent averages over successful replicas, with the number of replicas $N$, indicated. Cartesian dimensions are in nanometres. A) Forward translocation from the PTC to the exit. B) Reverse translocation from the exit to the PTC.}
    \label{fig:solvation}
\end{figure}

Motivated by these findings, we analysed how the solvation of the deca-alanine changes as it moves through the exit tunnel by computing, at each point along the translocation pathway, the number of water molecules located within the first solvation shell of the peptide. Fig.\,\ref{fig:solvation} shows the resulting trajectories, tracked using the position of the terminal residue of the peptide (the ACE cap in the forward direction, or the NME cap in the reverse direction), averaged over all successful translocation replicas and coloured according to the average number of surrounding water molecules.

Across all simulated systems and directions, the number of water molecules in the first solvation shell varies within a relatively narrow range, from approximately 55 to 85 molecules. However, the way this number evolves along the tunnel depends on the direction of translocation. In the forward direction, the peptide is markedly less solvated at the beginning of the trajectory, because it initially occupies the narrow region of the tunnel between the PTC and the CS, where the reduced luminal volume limits the number of water molecules that can pack around the chain. In the reverse direction, the opposite trend is observed: peptide solvation is highest at the start of the trajectory, reflecting the fact that the peptide initially resides in the wider vestibule near the tunnel exit, a region that can accommodate a substantially larger hydration shell. Indeed, the number of water molecules in this region is comparable to that around a free peptide (88$\pm$8), as calculated from the MD simulations used to generate the initial peptide conformations, indicating that the peptide near the tunnel exit is solvated to a similar extent as in bulk water.

Comparing the different tunnels considered in this study, the mitochondrial exit tunnel emerges as the most confined. In the forward direction, the peptide remains comparatively desolvated even after passing the constriction site, in contrast to the bacterial and especially cytosolic eukaryotic tunnels, where solvation increases more markedly beyond this point. This is consistent with the specialised architecture of the mitochondrial ribosome. Unlike cytosolic and bacterial ribosomes, which synthesise proteins of highly diverse sequence composition, mitochondrial ribosomes are dedicated to the production of a small set of highly hydrophobic proteins \cite{Gruschke2010b, Greber2014a}. The polypeptide tunnel exit of the mitochondrial ribosome has been shown to be structurally and compositionally tailored to this task.

Comparative structural analyses across the three domains of life further indicate that the lower part of the exit tunnel is substantially narrower and more occluded in eukaryotic and organellar ribosomes than in their bacterial counterparts \cite{DaoDuc2019}. A relatively less solvated, confined environment in the inner mitochondrial tunnel would therefore be functionally advantageous: it limits premature exposure of hydrophobic stretches of the nascent chain to the aqueous milieu, reducing the risk of misfolding or aggregation before the peptide reaches the membrane-insertion machinery, and is consistent with the general observation that hydrophobic side chains are disfavoured within the water-filled tunnel lumen relative to polar solvent \cite{Petrone2008}.

\section{Conclusions}

Using atomistic fpMD simulations, we assessed the intrinsic directionality of the ribosomal exit tunnel across four evolutionarily distinct ribosomes. Our main observation was that the reverse translocation failed considerably more often than translocation in the natural direction. Because the reverse-direction force profiles reported above are averaged only over the replicas that did reach the PTC, they necessarily disregard the trajectories in which the peptide got stuck. Had these unsuccessful replicas been included, the resistance associated with reverse translocation would appear even more pronounced than shown here.

These conclusions should be read alongside several limitations of our model. The tunnel was probed with a single homorepeat peptide, deca-alanine, and nascent chains of more realistic, heterogeneous sequence composition would likely interact with the tunnel walls quite differently, presumably more strongly than deca-alanine due to charged or bulky aromatic side chains. Furthermore, our tunnel models are cutouts of the full ribosome: harmonic restraints keep the surrounding rRNA and protein scaffold in place, so that fast dynamics are likely well captured, but larger-scale conformational changes of the tunnel that may occur \emph{in vivo} are precluded. Finally, because fpMD drives translocation many orders of magnitude faster than occurs during translation, part of the asymmetry we observe between forward and reverse translocation may reflect the kinetics of the pulling protocol rather than the underlying thermodynamics alone. The true Gibbs energy profile of the tunnel may therefore be less asymmetric than our force profiles suggest, and reverse translocation could be comparatively less obstructed at the slow, near-equilibrium rates relevant to translation \emph{in vivo}.

Still, if the tunnel architecture intrinsically disfavours reverse movement, this would explain why elements such as the mL45 N-terminal extension and the N-$\beta$NAC tail require dedicated structural machinery to achieve their transient reverse occupancy of the tunnel. Both proteins must act against, rather than with, the tunnel's intrinsic bias. Uncovering the detailed mechanistic basis of these processes will require more specialised simulations tailored to each system, although, at least in the NAC case, recent work has already begun to shed light on how it achieves its reverse insertion \cite{Lee2026}.

%\section*{Acknowledgment}

%text

\section*{Funding} 

The work was supported by the Czech Science Foundation (project 23-05557S). We acknowledge VSB--Technical University of Ostrava, IT4Innovations National Supercomputing Center, Czech Republic, for awarding this project access to the LUMI supercomputer, owned by the EuroHPC Joint Undertaking, hosted by CSC (Finland) and the LUMI consortium through the Ministry of Education, Youth and Sports of the Czech Republic through the e-INFRA CZ (grant ID: 90254).

\section*{Authors' contribution}

FCN setup, performed, and analysed MD simulations; PC built the simulation models of the tunnels and run initial peptide-free simulations; FCN, and MK interpreted the results; MK acquired funding; FCN, and MK wrote the initial draft of the manuscript; all authors finalized the manuscript.

\section*{Conflict of interest}

The authors declare no conflict of interest.

%\section*{Data and software availability}

%The input data necessary to run the MD simulations, the output data, and the scripts needed to reproduce all figures are provided at https://www.github.com/kolarlab/nepomuceno-tunnels. 

%\bibliography{refs-abbr}

\providecommand{\latin}[1]{#1}
\makeatletter
\providecommand{\doi}
  {\begingroup\let\do\@makeother\dospecials
  \catcode`\{=1 \catcode`\}=2 \doi@aux}
\providecommand{\doi@aux}[1]{\endgroup\texttt{#1}}
\makeatother
\providecommand*\mcitethebibliography{\thebibliography}
\csname @ifundefined\endcsname{endmcitethebibliography}
  {\let\endmcitethebibliography\endthebibliography}{}

\end{document}